\documentclass[aps,prl,twocolumn,showpacs,byrevtex]{revtex4}
\let\oldhat\hat
\renewcommand{\hat}[1]{\oldhat{\mathbf{#1}}}

\usepackage{times,mathptm}
\usepackage[dvips]{graphicx,color}
\usepackage{titlesec}

\begin{document}
\title{Two-dimensional topological semimetal states in monolayers Cu$_2$Ge, Fe$_2$Ge, and Fe$_2$Sn}
\author{Liangliang Liu$^{1,2}$, Chongze Wang$^{2}$, Jiangxu Li$^3$, Xing-Qiu Chen$^3$, Yu Jia$^{1\dagger}$, and Jun-Hyung Cho$^{2*}$}
\affiliation{$^1$ Key Laboratory for Special Functional Materials of Ministry of Education, Henan University, Kaifeng 475004, People's Republic of China \\
$^2$ Department of Physics, Research Institute for Natural Science, and HYU-HPSTAR-CIS High Pressure Research Center, Hanyang
University, 222 Wangsimni-ro, Seongdong-Ku, Seoul 04763, Republic of Korea \\
$^3$ Shenyang National Laboratory for Materials Science, Institute of Metal Research, Chinese Academy of Sciences, Shenyang 110016, China
}
\date{\today}

\begin{abstract}
Recent experimental realizations of the topological semimetal states in several monolayer systems are very attractive because of their exotic quantum phenomena and technological applications. Based on first-principles density-functional theory calculations including spin-orbit coupling, we here explore the drastically different two-dimensional (2D) topological semimetal states in three monolayers Cu$_2$Ge, Fe$_2$Ge, and Fe$_2$Sn, which are isostructural with a combination of the honeycomb Cu or Fe lattice and the triangular Ge or Sn lattice. We find that (i) the nonmagnetic (NM) Cu$_{2}$Ge monolayer having a planar geometry exhibits the massive Dirac nodal lines, (ii) the ferromagentic (FM) Fe$_2$Ge monolayer having a buckled geometry exhibits the massive Weyl points, and (iii) the FM Fe$_2$Sn monolayer having a planar geometry and an out-of-plane magnetic easy axis exhibits the massless Weyl nodal lines. It is therefore revealed that mirror symmetry cannot protect the four-fold degenerate Dirac nodal lines in the NM Cu$_{2}$Ge monolayer, but preserves the doubly degenerate Weyl nodal lines in the FM Fe$_{2}$Sn monolayer. Our findings demonstrate that the interplay of crystal symmetry, magnetic easy axis, and band topology is of importance for tailoring various 2D topological states in Cu$_2$Ge, Fe$_2$Ge, and Fe$_2$Sn monlayers.
\end{abstract}
\maketitle

\section{I. INTRODUCTION}

In the past decade, topological insulators and topological semimetals have attracted considerable attention because of their promising prospects in both fundamental research and technological applications~\cite{Wan,Bansil,Dai,Burk,Armitage}. Specifically, topological semimetals are characterized by the nontrivial topology of gapless bulk bands near the Fermi energy $E_{\rm F}$ and its associated robust surface states~\cite{Liu,Weng,Hasan}. There are several types of topological semimetals such as Dirac semimetal (DSM), Weyl semimetal (WSM), and nodal-line semimetal (NLS)~\cite{Yu,Yang,nodalLi,CaPfam}. The DSM (WSM) states have four-fold (two-fold) degenerate band crossings at discrete ${\bf k}$ points in momentum space, while the NLS states have band crossings along the closed or open lines within the Brillouin zone~\cite{HuangHQ,JHu,ELiu,Y2C,Burkov}. Interestingly, the NLS systems have drumhead-like surface states with narrow band dispersions, thereby giving rise to a high density of states near $E_{\rm F}$. As a result, such topologically nontrivial surface states are very vulnerable to various exotic phenomenona such as flatband ferromagnetism, Mott physics, high-$T_{\rm c}$ superconductivity, and other electronic instabilities~\cite{Bian,RLi,JHe}.

Most of the NLS states have so far been experimentally observed in three-dimensional (3D) materials such as PtSn$_4$~\cite{YWu}, ZrSiS~\cite{LSchoop}, and PbTaSe$_2$~\cite{GBian}. However, recent theoretical and experimental studies of such NLS states have been extended to 2D monolayers~\cite{BFeng,LGao,SJeon,BFeng2} whose electronic properties can be easily tuned by mechanical strains~\cite{hana,chow}. Based on the combined angle-resolved photoemission spectroscopy measurements and density-functional theory (DFT) calculations, Feng $et$ $al$.~\cite{BFeng} reported the presence of Dirac nodal lines (DNLs) in Cu$_{2}$Si monolayer which is composed of a honeycomb Cu lattice and a triangular Si lattice. Here, Cu$_{2}$Si monolayer has a planar geometry, which is identical to that of Cu$_{2}$Ge monolayer [see Fig. 1(a)]. Subsequently, Feng $et$ $al$.~\cite{BFeng2} also synthesized another isostructural monolayer of Ag$_{2}$Gd to observe Weyl nodal lines (WNLs) in the ferromagnetic (FM) phase. It is, however, noticeable that the DNLs of Cu$_{2}$Si monolayer and the WNLs of Ag$_{2}$Gd monolayer were predicted to lift their four-fold and two-fold degeneracies with including spin-orbit coupling (SOC)~\cite{BFeng,BFeng2}, respectively. Therefore, Cu$_{2}$Si and Ag$_{2}$Gd monolayers have non-zero masses in DNLs and WNLs, respectively. We note that the nonmagnetic (NM) phase of Cu$_{2}$Si monolayer preserves the mirror symmetry, while the FM phase of Ag$_{2}$Gd monolayer having the in-plane magnetic easy axis breaks the mirror symmetry ~\cite{ACorrea,MOrmaza}. However, if the magnetization direction in the latter monolayer were reoriented along the out-of-plane direction via external perturbations, e.g., spin-orbit torque, the mirror symmetry would be respected to protect massless WNLs against SOC, as demonstrated below in Fe$_2$Sn monolayer. Thus, the symmetry protection of WNLs in FM monolayers can be manipulated by the spin reorientation effect~\cite{QWang,JJin,RWang1}.

\begin{figure*}[ht]
\centering{ \includegraphics[width=16.cm]{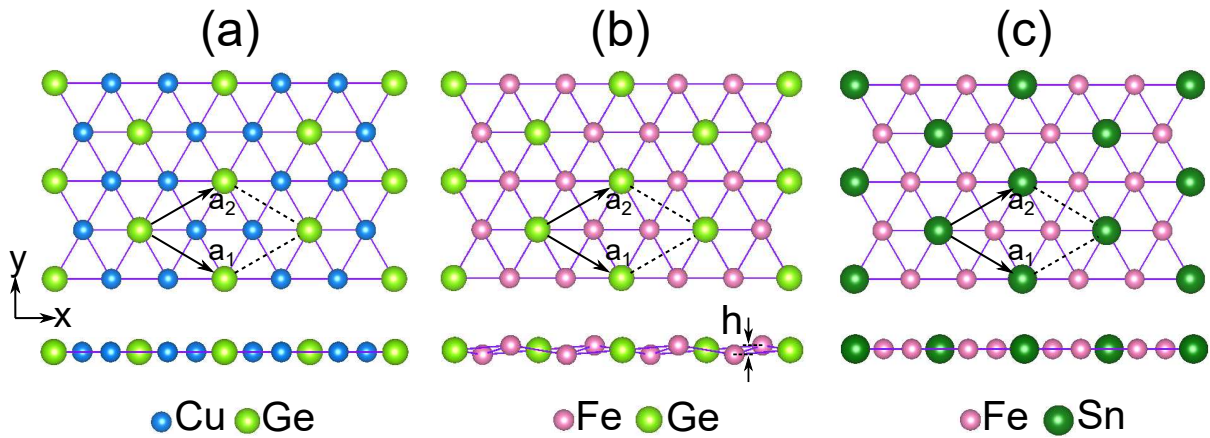} }
\caption{(Color online) Optimized structures of (a) Cu$_{2}$Ge, (b) Fe$_{2}$Ge, and (c) Fe$_{2}$Sn monolayers. The top and side views are given in the upper and lower panels, respectively. The arrows represent the primitive lattice vectors $a_1$ and $a_2$ in each unit cell (indicated by the dashed lines). The structures of Cu$_{2}$Ge and Fe$_{2}$Sn monolayers are planar, while that of Fe$_{2}$Ge monolayer is buckled with a height difference $h$ between neighboring Fe atoms.}
\end{figure*}

In this paper, we systematically investigate the different 2D topological states in three monolayers Cu$_{2}$Ge, Fe$_{2}$Ge, and Fe$_{2}$Sn using DFT calculations with the inclusion of SOC. We find that these monolayers have different ground states depending on the presence/absence of magnetism and mirror symmetry: i.e., massive DNLs for the NM Cu$_2$Ge monolayer having the mirror symmetry with respect to the $x$-$y$ plane [see Fig. 1(a)], massive Weyl points for the FM Fe$_{2}$Ge monolayer breaking the mirror symmetry [Fig. 1(b)], and massless WNLs for the FM Fe$_2$Sn monolayer having the mirror symmetry with an out-of-plane magnetic easy axis [Fig. 1(c)]. It is thus revealed that for the NM Cu$_{2}$Ge monolayer, the mirror symmetry cannot protect the four-fold degeneracy of DNLs, whereas for the FM Fe$_{2}$Sn monolayer, it protects the two-fold degeneracy of WNLs. Interestingly, unlike Fe$_2$Sn monolayer, the geometry of Fe$_{2}$Ge monolayer is found to be buckled due to an increased magnetic stress arising from its relatively smaller lattice constants. The resulting broken mirror symmetry in Fe$_{2}$Ge monolayer induces a transformation from the WNLs to Weyl points. Therefore, our comprehensive investigation of different 2D topological quantum states in Cu$_2$Ge, Fe$_2$Ge, and Fe$_2$Sn monolayers demonstrates that the versatile topological behaviors can be entangled with mirror symmetry and magnetism.

\section{II. Calculational methods}

The present DFT calculations were performed using the Vienna {\it ab initio} simulation package with the projector-augmented wave method~\cite{vasp1,vasp2,paw}. For the exchange-correlation energy, we employed the generalized-gradient approximation functional of Perdew-Burke-Ernzerhof (PBE)~\cite{pbe}. The present monoalyer systems were modeled by a periodic slab geometry with ${\sim}$30 {\AA} of vacuum in between the slabs. A plane-wave basis was employed with a kinetic energy cutoff of 500 eV, and the ${\bf k}$-space integration was done with the 21${\times}$21 meshes in the 2D Brillouin zone. All atoms were allowed to relax along the calculated forces until all the residual force components were less than 0.005 eV/{\AA}. To investigate the topological properties of Cu$_2$Ge, Fe$_2$Ge and Fe$_2$Sn monolayers, we constructed Wannier functions by projecting the Bloch electronic states obtained from DFT calculations onto a set of Cu (Fe) $s$, Cu (Fe) $d$, and Ge (Sn) $p$ orbitals. Based on the tight-binding Hamiltonian with a basis of maximally localized Wannier functions~\cite{wannier90}, we not only identified the existence of nodal lines but also calculated the Berry curvature around the band crossing points by using the WANNIERTOOLS package~\cite{wanniertools}.

\section{III. Results}

We begin by optimizing the atomic structures of Cu$_{2}$Ge, Fe$_{2}$Ge, and Fe$_{2}$Sn monolayers, where Cu or Fe (Ge or Sn) atoms form a honeycomb (triangular) lattice. Their optimized structures are displayed in Figs. 1(a), 1(b), and 1(c), respectively. For Cu$_2$Ge monolayer, we obtain the lattice constants $a_1$ = $a_2$ = 4.218 {\AA} with a planar geometry, in good agreement with those ($a_1$ = $a_2$ = 4.214 {\AA}) of a previous DFT calculation~\cite{Cu2Gemono}. Therefore, Cu$_{2}$Ge monolayer has the point group of $D_{\rm 6h}$ with the mirror symmetry $M_{\rm z}$ about the $x$-$y$ plane. Meanwhile, Fe$_{2}$Ge monolayer is found to be buckled with $a_1$ = $a_2$ = 4.147 {\AA} and a height difference of 0.403 \AA\ between neighboring Fe atoms [see Fig. 1(b)]. This buckling of Fe atoms is induced by the emergence of FM order, as discussed below. Therefore, Fe$_{2}$Ge monolayer has the broken $M_{\rm z}$ mirror symmetry, leading to a reduced crystalline point group $D_{\rm 3d}$. Meanwhile, the FM Fe$_{2}$Sn monolayer has a planar geometry with $a_1$ = $a_2$ = 4.453 {\AA}, preserving the crystalline point group of $D_{\rm 6h}$ with the $M_{\rm z}$ mirror symmetry. We note that the equilibrium structures of Cu$_{2}$Ge, Fe$_{2}$Ge, and Fe$_{2}$Sn monolayers do not exhibit any imaginary phonon mode in their calculated phonon dispersions, indicating that they are thermodynamically stable (see Fig. S1 in the Supplemental Material~\cite{supple}).

Figure 2(a) shows the electronic band structure of Cu$_{2}$Ge monolayer in the absence of SOC. We find that the two hole-like bands (labeled as $\alpha$ and $\beta$) and one electron-like band ($\gamma$) overlap with each other near the Fermi level $E_{\rm F}$. As a result, there are four nodal points [designated as A, A$^\prime$, B, and B$^\prime$ in Fig. 2(a)] along the ${\Gamma}-$M and ${\Gamma}-$K lines. Using the tight-binding Hamiltonian with a basis of maximally localized Wannier functions~\cite{wannier90,wanniertools}, we reveal the existence of two DNLs around the ${\Gamma}$ point [see Fig. 2(b)]. As shown in Fig. S2 in the Supplemental Material~\cite{supple}, the Wannier bands near $E_{\rm F}$ are in good agreement with
the DFT bands. It is noticeable that the two DNLs are protected by the $M_{z}$ mirror symmetry with two different one-dimensional irreducible symmetry representations: i.e., if the two crossing bands have the opposite eigenvalues of $M_{z}$, they cannot hybridize with each other~\cite{nodaline,nodaline2}. The band projections onto the Cu 3$d$ and Ge 4$p$ orbitals show that the $\alpha$ and $\beta$ bands are mainly composed of the Cu \textit{d}$_{xy}$/\textit{d}$_{x^{2}-y^{2}}$ and Ge \textit{p}$_{x}$/\textit{p}$_{y}$ orbitals, while the $\gamma$ band arises from the Cu \textit{d$_{xz}$} and Ge \textit{p}$_{z}$ orbitals (see Fig. S3 in the Supplemental Material~\cite{supple}). Therefore, the $\alpha$ and $\beta$ bands have the even parity eigenvalue of $M_{z}$, which is opposite to the odd parity eigenvalue for the $\gamma$ band [see Fig. 2(a)]. We note that, when $M_{z}$ is broken by the buckling of two Cu atoms in the primitive unit cell, the DNL containing A and A$^\prime$ (B and B$^\prime$) is transformed into three Dirac points along the three nonequivalent ${\Gamma}-$K (${\Gamma}-$M) lines (see Fig. S4 in the Supplemental Material~\cite{supple}).

\begin{figure}[h!t]
\centering{ \includegraphics[width=8 cm]{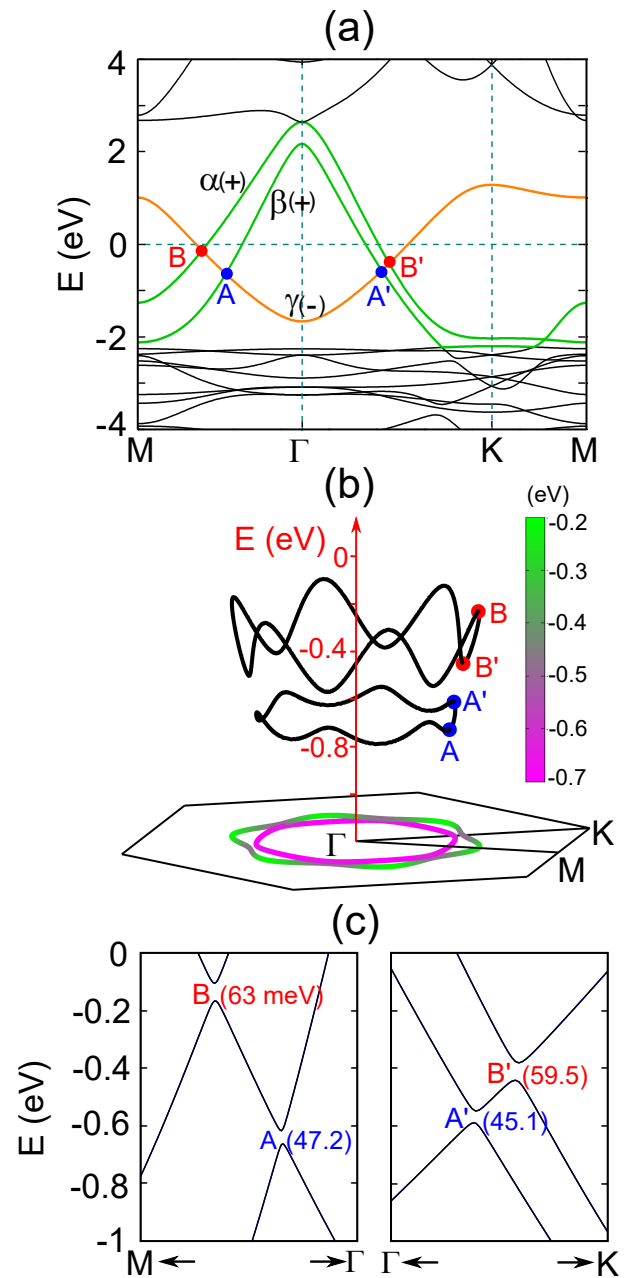} }
\caption{(Color online) (a) Calculated band structure of Cu$_{2}$Ge monolayer in the absence of SOC. The four crossing points of three bands (labeled as $\alpha$, $\beta$, and $\gamma$) along the ${\Gamma}-$M and ${\Gamma}-$K lines are designated as A, A$^{\prime}$, B, and B$^{\prime}$. The ${\Gamma}-$M direction is parallel to the $x$ axis. For the $\alpha$, $\beta$, and $\gamma$ bands, the parity of mirror symmetry is labeled plus or minus sign in parentheses. (b) Energy dispersions of the two DNLs passing through the A and A$^{\prime}$ (B and B$^{\prime}$) points, together with their projections onto the Brillouin zone using the color scale. (c) Zoom-in band structures around the A and B (A$^{\prime}$ and B$^{\prime}$) points, obtained with including SOC. The numbers represent the gaps (in meV) at the A, A$^{\prime}$, B, and B$^{\prime}$ points.}
\end{figure}

In order to examine how SOC influences the four-fold degeneracy of DNLs in Cu$_{2}$Ge monolayer, we perform the DFT calculations with including SOC. Figure 2(c) displays the calculated band structures along the ${\Gamma}-$M and ${\Gamma}-$K lines around the DNLs. We find the band-gap openings of about 45$-$63 meV at the crossing points of DNLs. However, each band preserves a two-fold degeneracy with the opposite parity eigenvalues $\pm i$ of $M_{z}$. It is noted that along the DNLs with the four-fold degeneracy, the two bands having the same parity eigenvalue can hybridize with each other, thereby opening band gaps. Therefore, $M_{z}$ in Cu$_{2}$Ge monolayer can not protect the DNLs against SOC, leading to the formation of massive DNLs.

\begin{figure}[ht]
\centering{ \includegraphics[width=8.cm]{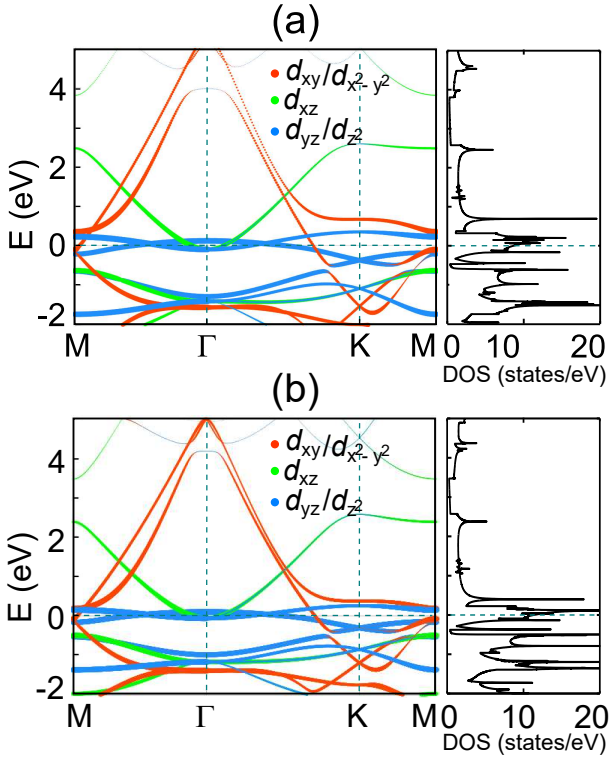} }
\caption{(Color online) Calculated band structures and DOS of the NM (a) Fe$_2$Ge and (b) Fe$_2$Sn monolayers. The band projections onto the Fe 3$d$ orbitals are displayed with circles whose radii are proportional to the weights of the $\textit{d$_{xy}$}$/$\textit{d$_{x^2-y^2}$}$, $\textit{d$_{xz}$}$, and $\textit{d$_{z^2}$}$/$\textit{d$_{yz}$}$ orbitals.}
\end{figure}

To realize the symmetry-protected nodal lines in 2D monolayers, it is prerequisite to split two-fold degenerate bands via the emergence of ferromagnetism that breaks time reverse symmetry. In the present study, we consider the two FM Fe$_2$Ge and Fe$_2$Sn monolayers. For Fe$_{2}$Ge and Fe$_2$Sn monolayers, the FM phase is found to be more stable than the NM (antiferromagnetic) one by 0.758 (0.404) eV and 1.181 (0.413) eV per unit cell, respectively. The calculated magnetic moments for the FM Fe$_2$Ge and Fe$_2$Sn monolayers are \textit{m} = 2.08 and 2.38 ${\mu}_{\rm B}$ per Fe atom, respectively. Figures 1(b) and 1(c) display the optimized structures of the FM Fe$_{2}$Ge and Fe$_2$Sn monolayers, respectively. Interestingly, for Fe$_{2}$Ge monolayer, the FM structure is buckled, while the NM one is planar. Here, the lattice constants ($a_1$ = $a_2$ = 4.147 {\AA}) of the FM structure are larger than those ($a_1$ = $a_2$ = 4.063 {\AA}) of the NM one (see Table S1 in the Supplemental Materials~\cite{supple}). Such buckling of Fe atoms and larger lattice constants in the FM phase can be attributed to the magnetic stress generated by the exchange interactions~\cite{press} of spin-polarized electrons. By contrast, the FM Fe$_{2}$Sn monolayer having the relatively larger lattice constants ($a_1$ = $a_2$ = 4.453 {\AA}) exhibits a planar geometry [see Fig. 1(c)].

\begin{figure*}[ht]
\centering{ \includegraphics[width=16.cm]{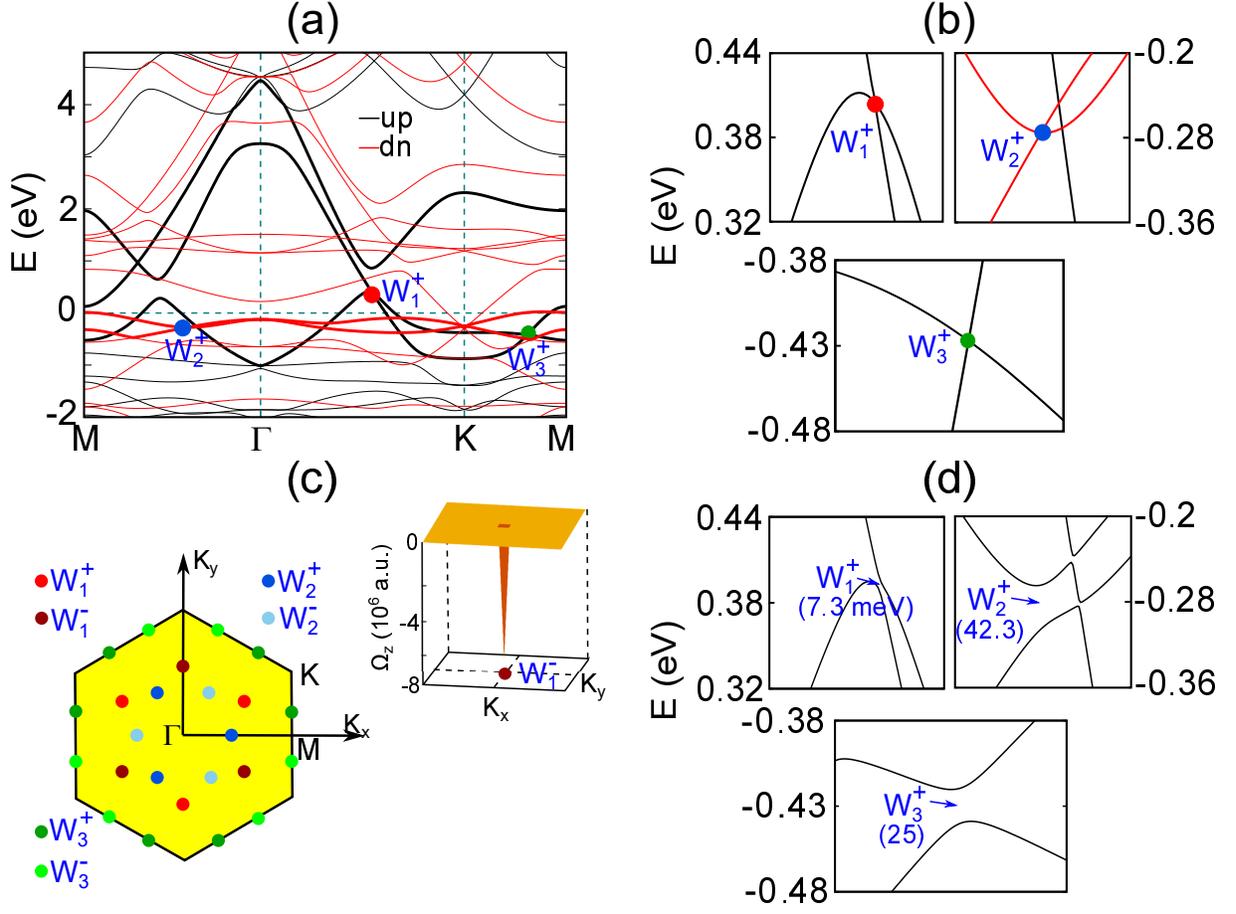} }
\caption{(Color online) (a) Calculated band structure of the FM Fe$_{2}$Ge monolayer in the absence of SOC and (b) zoom-in band structures around the three Wely points W$^{+}_{1}$, W$^{+}_{2}$, and W$^{+}_{3}$. (c) Distribution of all the Weyl points in the whole Brillouin zone, together with the Berry curvature component $\Omega_{z}$ around W$^{-}_{1}$. (d) Zoom-in band structures around the W$^{+}_{1}$, W$^{+}_{2}$, and W$^{+}_{3}$ points in (b), obtained with including SOC. The numbers represent the gaps (in meV) at W$^{+}_{1}$, W$^{+}_{2}$, and W$^{+}_{3}$.}
\end{figure*}

Next, we examine why the FM instability exists in Fe$_2$Ge and Fe$_2$Sn monolayers. For this, we calculate the band structure and density of states (DOS) for their NM phases. The calculated NM band structure of Fe$_2$Ge monolayer [see Fig. 3(a)] is similar to that [Fig. 3(b)] of Fe$_2$Sn monolayer, showing that the electronic states around $E_{\rm F}$ are mostly composed of the Fe 3\textit{d} orbitals (see also Fig. S5 in the Supplemental Material~\cite{supple}). We find that above $E_{\rm F}$, the two hole-like and one electron-like parabolic bands arise from the Fe \textit{d$_{xy}$}, \textit{d$_{x^2-y^2}$}, and \textit{d$_{xz}$} orbitals with effectively high neighbor hoppings, giving rise to large energy dispersions [see Figs. 3(a) and 3(b)]. Meanwhile, close to $E_{\rm F}$, there exist the flatbands arising from the Fe \textit{d$_{z^2}$} and \textit{d$_{yz}$} orbitals, which are relatively more localized than the \textit{d$_{xy}$}, \textit{d$_{x^2-y^2}$}, and \textit{d$_{xz}$} components. Here, unlike the \textit{d$_{xz}$} orbital, the \textit{d$_{yz}$} orbital is deviated away from the Fe$-$Fe bonds directing parallel to the $x$ axis. It is noted that the 3\textit{d$^6$}4\textit{s$^2$} valence electrons of Fe atom are less than those (3\textit{d$^{10}$}4\textit{s$^1$}) of Cu atom. Therefore, the positions of $E_{\rm F}$ in Fe$_{2}$Ge and Fe$_{2}$Sn monolayers [see Figs. 3(a) and 3(b)] shift downward relative to that in Cu$_{2}$Ge monolayer [Fig. 2(a)]. As a consequence of such Fermi level shifts, Fe$_{2}$Ge and Fe$_{2}$Sn monolayers have high DOS at $E_{\rm F}$, which in turn induces a FM order via the Stoner criterion $D(E_{\rm{F}})I > 1$~\cite{stoner1,stoner2} (see Fig. S6 in the Supplemental Material~\cite{supple}). Here, $D(E_{\rm{F}})$ is the total DOS at $E_{\rm F}$ and the Stoner parameter $I$ can be estimated with dividing the exchange splitting of spin-up and spin-down bands by the corresponding magnetic moment.

Figure 4(a) shows the band structure of the FM phase of Fe$_{2}$Ge monolayer, computed without including SOC. We find that near $E_{\rm F}$, there are three spinful Weyl points W$_{1}^{+}$, W$_{2}^{+}$, and W$_{3}^{+}$ along the ${\Gamma}-$M, ${\Gamma}-$K, and K$-$M lines, respectively [see Fig. 4(b) and Fig. S7 in the Supplemental Material~\cite{supple}]. Here, we consider the crossings of the same spin-polarized bands because the absence of SOC decouples two different spin channels. It is noted that the crystalline point group $D_{3d}$ of buckled Fe$_{2}$Ge monolayer has three generators including threefold rotational symmetry $C_{3z}$ about the $z$ axis, inversion symmetry $P$, and mirror symmetry $M_{y}$ about the $x$-$z$ plane. Therefore, we have not only nine nonequivalent Weyl points of W$_{1}^{+}$, W$_{2}^{+}$, and W$_{3}^{+}$ in the whole Brillouin zone, but also their paired Weyl points W$_{1}^{-}$, W$_{2}^{-}$, and W$_{3}^{-}$ of opposite chirality at inversion symmetric ${\bf k}$-points [see Fig. 4(c)]. It has been known that specific crystalline symmetries are needed to guarantee 2D massless Weyl points~\cite{2Dweyl,2Dweyl2}. For Fe$_{2}$Ge monolayer, the twofold degeneracy of Weyl points in the ${\Gamma}-$K and K$-$M lines is mandated by $C_2$ (equivalent to the combination of $P$ and $M_{y}$), because the two crossing bands have the opposite parity eigenvalues ${\pm}1$ of $C_2$. Meanwhile, the Weyl points in the ${\Gamma}-$M line are protected by $M_{y}$ (see Fig. S8 in the Supplemental Material~\cite{supple}). Using the WANNIERTOOLS package~\cite{wanniertools}, we demonstrate that each pair of Weyl points have the positive and negative Berry curvature distributions [see Fig. 4(c)]. However, the inclusion of SOC lifts the two-fold degeneracy of all Weyl points, leading to massive Weyl points. Figure 4(d) displays the gap openings of W$_{1}^{+}$, W$_{2}^{+}$, and W$_{3}^{+}$. It is noteworthy that SOC aligns the spontaneous magnetization direction parallel to the $z$ axis, as discussed below. Such a magnetic anisotropy breaks $C_2$ and $M_{y}$ symmetries, thereby giving rise to the SOC-induced gap opening at each Weyl point.

\begin{figure*}[ht]
\centering{ \includegraphics[width=17.cm]{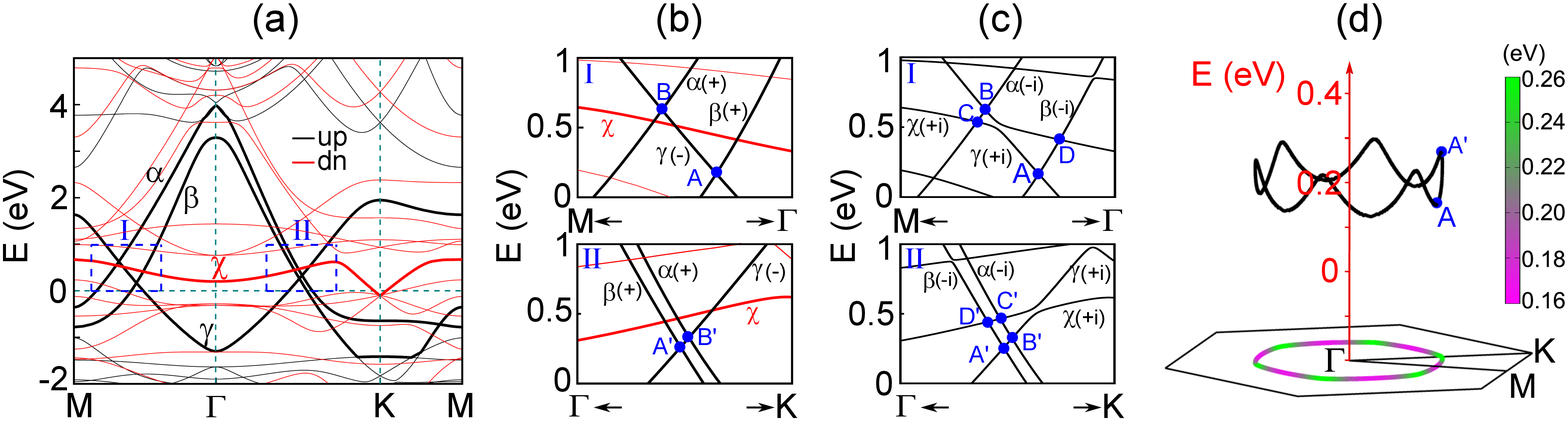} }
\caption{(Color online)  (a) Calculated spin-polarized band structure of Fe$_2$Sn monolayer in the absence of SOC. The three spin-up and one spin-down bands around $E_{\rm F}$ are labeled as $\alpha$, $\beta$, $\gamma$, and $\chi$ (thick red line), respectively. Zoom-in band structures in the dashed square are shown in (b), where the crossing points of the $\alpha$, $\beta$, and $\gamma$ bands are designated as A, B, A$^{\prime}$ and B$^{\prime}$. The parity of mirror symmetry is labeled plus or minus sign in the $\alpha$, $\beta$, and $\gamma$ bands. The corresponding zoom-in band structures in the presence of SOC are given in (c), where ${\pm}i$ represents the parity of mirror symmetry. (d) Energy dispersion of the WNL passing through the A and A$^{\prime}$ points in (c).}
\end{figure*}

Contrasting with the buckled geometry of Fe$_{2}$Ge monolayer, Fe$_{2}$Sn monolayer has a planar geometry which involves the same crystalline point group of $D_{\rm 6h}$ as Cu$_2$Ge monolayer. The resulting preservation of mirror symmetry $M_{z}$ in Fe$_{2}$Sn monolayer will be demonstrated to allow the protection of WNLs in the presence of SOC. Figure 5(a) shows the calculated band structure of the FM Fe$_{2}$Sn monolayer without SOC. We find that the spin-up bands exhibit the overlaps of two hole-like (labeled as $\alpha$ and $\beta$) and one electron-like ($\gamma$) bands around $E_{\rm F}$, giving rise to four crossing points [designated as A, A$^\prime$, B, and B$^\prime$ in Fig. 5(b)] along the ${\Gamma}-$M and ${\Gamma}-$K lines. These nodal points evolve into the two WNLs in the whole Brillouin zone: one passes through the A and A$^{\prime}$ points and the other passes through the B and B$^{\prime}$ points (see Fig. S9 in the Supplemental Material). The band projections onto the Fe 3$d$ orbitals show that the $\alpha$ ($\beta$) band is mainly composed of the Fe \textit{d}$_{x^{2}-y^{2}}$ (\textit{d}$_{xy}$) orbital with the even parity of $M_z$, while the $\gamma$ band arises from the Fe \textit{d}$_{xz}$ orbital with the odd parity of $M_z$ (see Fig. S10 in the Supplemental Material~\cite{supple}). Therefore, the two-fold degeneracy of the two WNLs is respected by $M_z$. When SOC is included, the spontaneous magnetization direction is also parallel to the $z$ axis, as discussed below. Therefore, the point group becomes C$_{6h}$ containing $M_{z}$. Figure 5(c) displays the SOC-included band structure of Fe$_{2}$Sn monolayer along the $\Gamma$-M and $\Gamma$-K lines around the crossing points. Since spin is not a good quantum number in the presence of SOC, the spin-up bands could hybridize with the spin-down ones. As shown in Fig. 5(c), the original crossing points (A, A$^\prime$, B, and B$^\prime$) are still reserved, but some additional crossing points (C, C$^{\prime}$, D, and D$^{\prime}$) appear because a spin-down band (labeled as $\chi$) overlaps with the spin-up $\alpha$, $\beta$ and $\gamma$ bands. Such nodal points evolve into the four WNLs passing through the A$-$A$^{\prime}$, C$-$B$^{\prime}$, B$-$C$^{\prime}$, and D$-$D$^{\prime}$ points, respectively (see Fig. S11 in the Supplemental Material~\cite{supple}). Here, the three WNLs passing through the C$-$B$^{\prime}$, B$-$C$^{\prime}$, and D$-$D$^{\prime}$ points are newly formed by the hybridization of spin-up and spin-down bands through SOC. Figure 5(d) shows the energy dispersion of the WNL passing through the A and A$^{\prime}$ points, which has a bandwidth of $\sim$100 meV. The other WNLs have relatively larger bandwidths (see Fig. S11 in the Supplemental Material~\cite{supple}). It is noted that each WNL is composed of two bands of the opposite $M_{z}$ parity eigenvalues $\pm$\textit{i} (see Fig. S10 in the Supplemental Material~\cite{supple}). Therefore, we can say that the predicted four WNLs in Fe$_2$Sn monolayer are robust against breaking the $M_{z}$ symmetry.

\begin{table}[ht]
\caption{Calculated MAE values (in $\mu$eV per Fe atom) of Fe$_{2}$Ge and Fe$_{2}$Sn monolayers with respect to the magnetic easy axis in the out-of-plane direction.}
\begin{ruledtabular}
\begin{tabular}{lccccc}
 	    & ${\Delta}E$[100]  &  ${\Delta}E$[110]  	 & ${\Delta}E$[3.732,1,0]  	 &  ${\Delta}E$[111] \\     \hline
Fe$_{2}$Ge     & 605    &  605 & 605	& 403			\\
Fe$_{2}$Sn     &   890  &  890 & 890    & 593 \\

 \end{tabular}
\end{ruledtabular}
\end{table}

Finally, we discuss magnetocrystalline anisotropic energy (MAE) which determines the orientation of magnetization as well as topological property. Table I shows the relative energy differences of Fe$_{2}$Ge and Fe$_{2}$Sn monolayers, depending on different magnetization directions. We find that the magnetic easy axes of both Fe$_{2}$Ge and Fe$_{2}$Sn monolayers are parallel to the out-of-plane direction. The calculated MAE values of Fe$_{2}$Ge (Fe$_{2}$Sn) monolayer are 605 (890) and 403 (593) $\mu$eV per Fe atom along the [100] and [111] directions, respectively. It is noted that the MAE values along other in-plane directions such as the [110] and [3.732,1,0] directions are the same as that along the [100] direction (see Table I). This invariance of MAE in the $x$-$y$ plane is likely to be due to the fact that the FM instability is induced by the highly localized Fe $d_{z^2}$ and $d_{yz}$ orbitals near $E_{\rm F}$ [see Figs. 3(a) and 3(b)]. Interestingly, the magnitudes of MAE in Fe$_{2}$Ge and Fe$_{2}$Sn monolayers are much larger than those of the typical FM crystals such as Fe (${\sim}$2 $\mu$eV), Co (${\sim}$65 $\mu$eV), and Ni (${\sim}$3 $\mu$eV)~\cite{FeCoNi,FeCoNi2}, as well as ${\sim}$400 $\mu$eV of the previously predicted 2D nodal-line materials InC~\cite{SJeon} and MnN~\cite{loopMnN}. Therefore, Fe$_{2}$Ge and Fe$_{2}$Sn monolayers can be classified as hard magnetic 2D materials, which barely change the spin orientations via externally applied magnetic field. It is noteworthy that Cu$_2$Si monolayer was synthesized by the deposition of Si atoms on the Cu(111) surface~\cite{BFeng} and Ag$_2$Gd monolayer was also synthesized by the deposition of Gd atoms on the Ag (111) surface~\cite{BFeng2}. In this sense, we anticipate that Cu$_{2}$Ge, Fe$_{2}$Ge, and Fe$_{2}$Sn monolayers could be synthesized by using the atomic layer deposition technique~\cite{ALD} in future experiments.

\section{IV. Summary}

We have performed first-principles calculations for Cu$_{2}$Ge, Fe$_{2}$Ge and Fe$_{2}$Sn monolayers to investigate their different 2D topological states. By a systematic study of the electronic structures of Cu$_2$Ge, Fe$_{2}$Ge, and Fe$_{2}$Sn monolayers, we revealed the existence of massive DNLs, massive Weyl points, and massless WNLs, respectively. Such different topological states were identified to be formed depending on the crystalline symmetries and NM/FM orders in the three monolayers, i.e., the planar NM Cu$_2$Ge monolayer with the mirror symmetry of $M_z$, the buckled FM Fe$_{2}$Ge monolayer breaking $M_z$, and the planar FM Fe$_{2}$Sn monolayer preserving $M_z$. Therefore, for Cu$_{2}$Ge monolayer, the mirror symmetry cannot protect the four-fold degeneracy of DNLs, but for Fe$_{2}$Sn monolayer, it protects the two-fold degeneracy of WNLs. Specifically, Fe$_{2}$Ge and Fe$_2$Sn monolayers have the sizable MAE values of one or two orders larger than those of Fe, Co, and Ni crystals. The resulting topological and magnetic properties of Fe$_{2}$Ge and Fe$_2$Sn monolayers are anticipated to be very promising for the utilization of future spintronics applications. Our findings demonstrated that the versatile topological properties can be entangled with mirror symmetry and time reversal symmetry in atomically thin monolayer systems.

This work was supported by the National Research Foundation of Korea (NRF) Grant funded by the
Korean Government (Grants No. 2019R1A2C1002975, No. 2016K1A4A3914691, and No. 2015M3D1A1070609), the NSFC (Grant No. 11774078), and the Innovation Scientists and Technicians Troop Construction Projects of Henan Province (Grant No. 10094100510025). The calculations were performed by the KISTI Supercomputing Center through the Strategic Support Program (Program No.
KSC-2018-CRE-0063) for the supercomputing application research and by the High Performance Computational Center
of Henan University.

Corresponding authors: $^{*}$chojh@hanyang.ac.kr, $^{\dagger}$jiayu@zzu.edu.cn



\newpage

\onecolumngrid
\newpage
\titleformat*{\section}{\LARGE\bfseries}

\renewcommand{\thefigure}{S\arabic{figure}}
\setcounter{figure}{0}

\vspace{1.2cm}

\section{Supplemental Material for “Two-dimensional topological semimetal states in monolayers Cu$_2$Ge, Fe$_2$Ge, and Fe$_2$Sn”}
\vspace{1.2cm}
\begin{flushleft}

{\bf 1. Calculated phonon spectra of Cu$_2$Ge, Fe$_2$Ge and Fe$_2$Sn monolayers.}
\begin{figure}[ht]
\includegraphics[width=11cm]{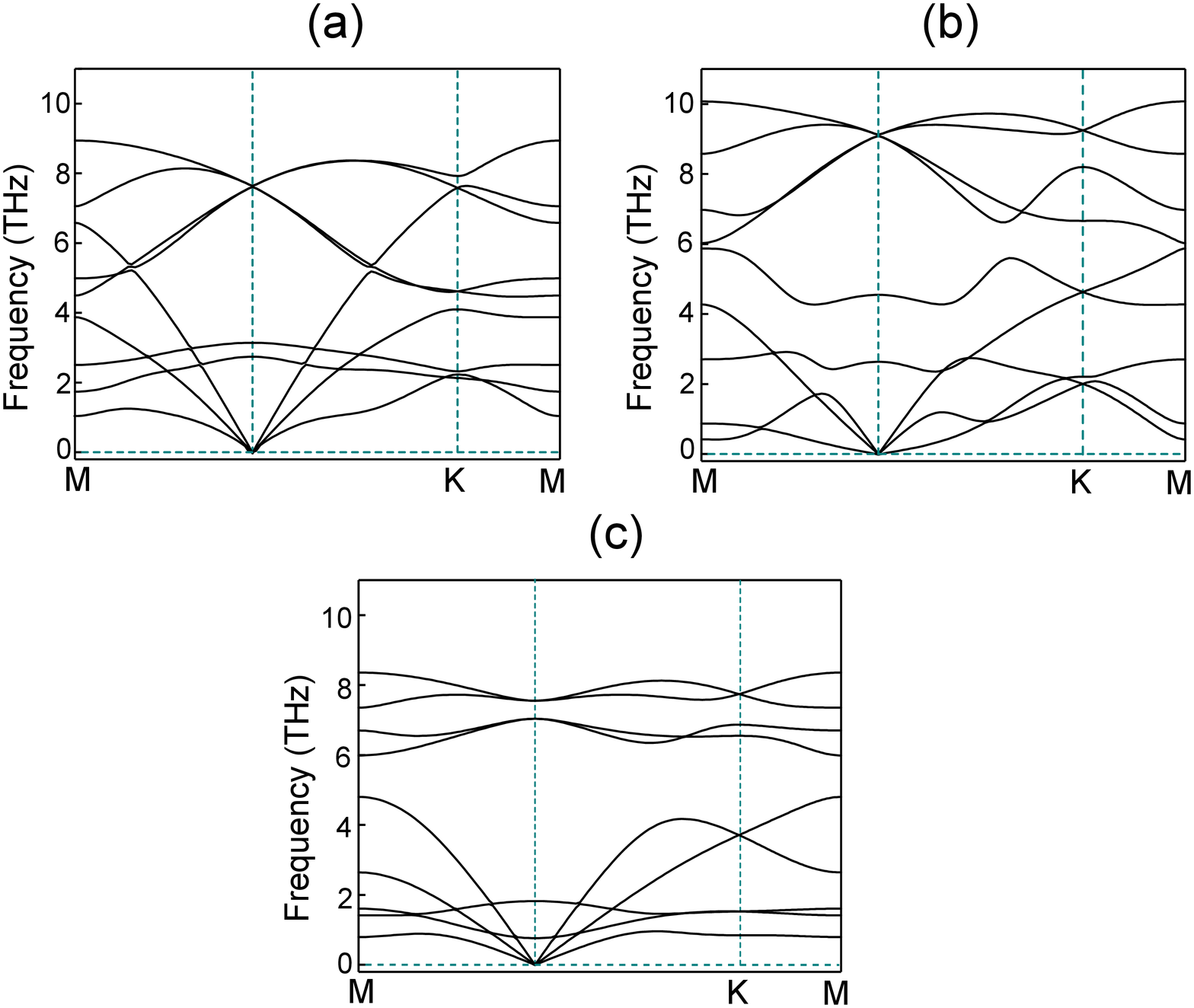}
\caption{ Calculated phonon spectra of the ground-state structures of (a) Cu$_2$Ge, (b) Fe$_2$Ge, and (c) Fe$_2$Sn monolayers using the Phonopy code~\cite{Phononpy}. The three equilibrium structures are found to be stable without any imaginary phonon mode. }
\end{figure}

\vspace{1.2cm}

{\bf 2. Electronic band structure of Cu$_2$Ge monolayer.}
\begin{figure}[ht]
\includegraphics[width=6.5cm]{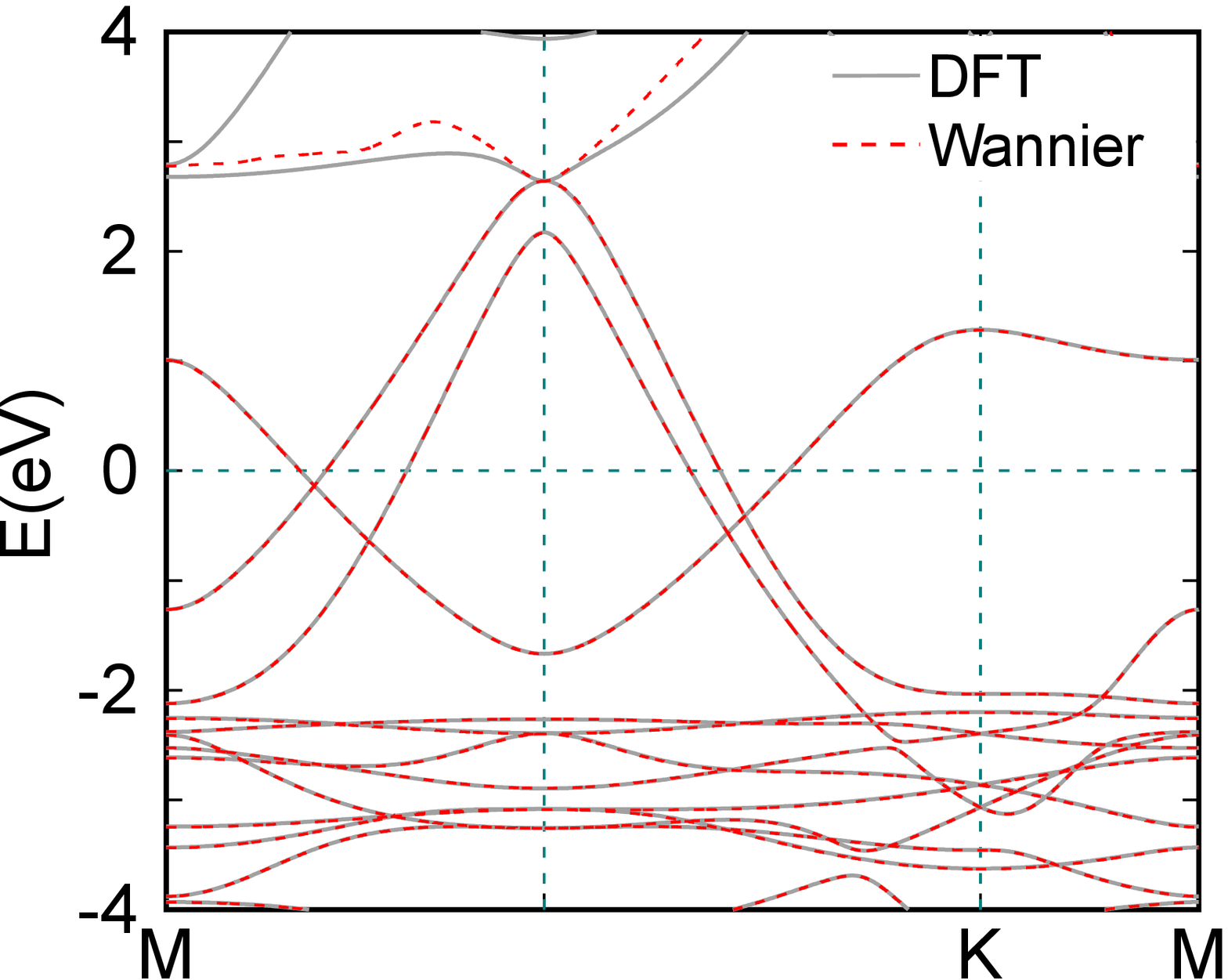}
\caption{ Band structure of Cu$_2$Ge monolayer, obtained using the tight-binding Hamiltonian with maximally localized Wannier functions (MLWF)~\cite{wannier}. The Wannier bands near the Fermi energy fit well with the DFT bands. }
\end{figure}


{\bf 3. Band projections onto Cu and Ge orbitals.}
\begin{figure}[ht]
\includegraphics[width=14.cm]{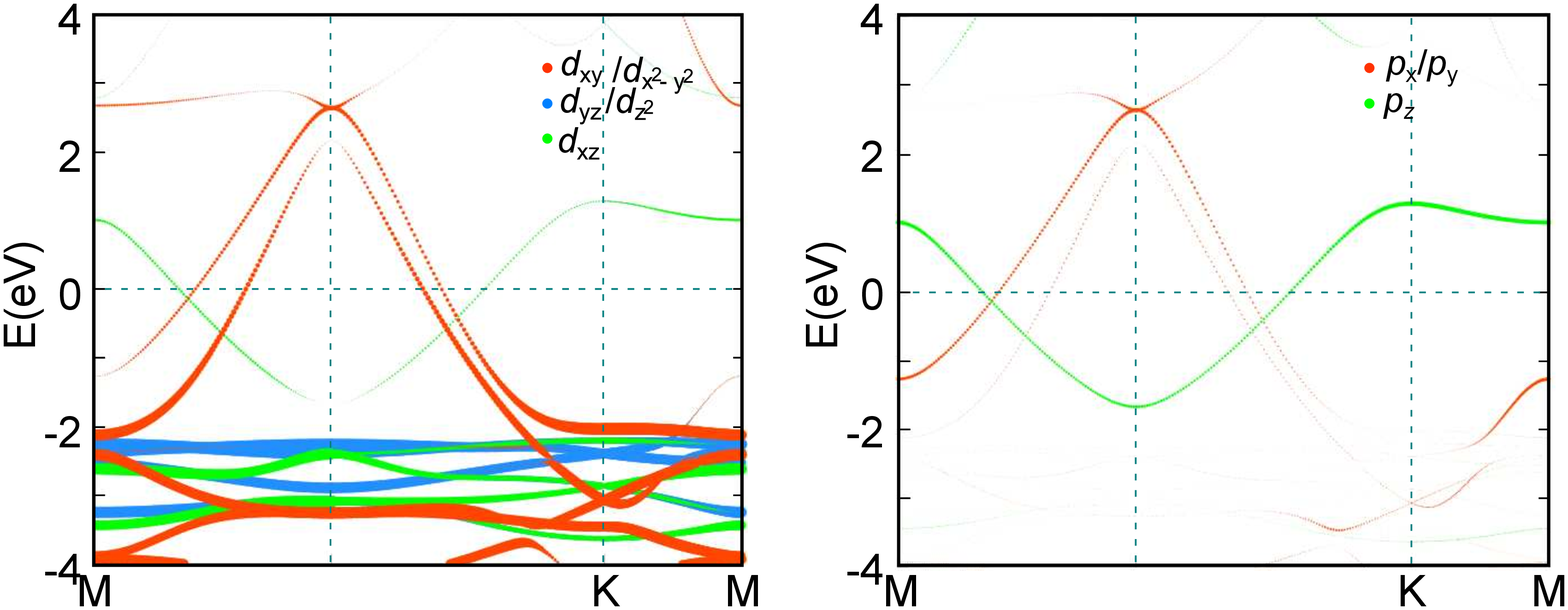}
\caption{ Calculated projected bands of Cu$_2$Ge monolayer onto the Cu 3$d$ and Ge 4$p$ orbitals. Here, the radii of circles are proportional to the weights of the corresponding orbitals.}
\end{figure}

\vspace{1.2cm}

{\bf 4. Electronic structure of buckled Cu$_2$Ge monolayer.}
\begin{figure}[ht]
\includegraphics[width=8cm]{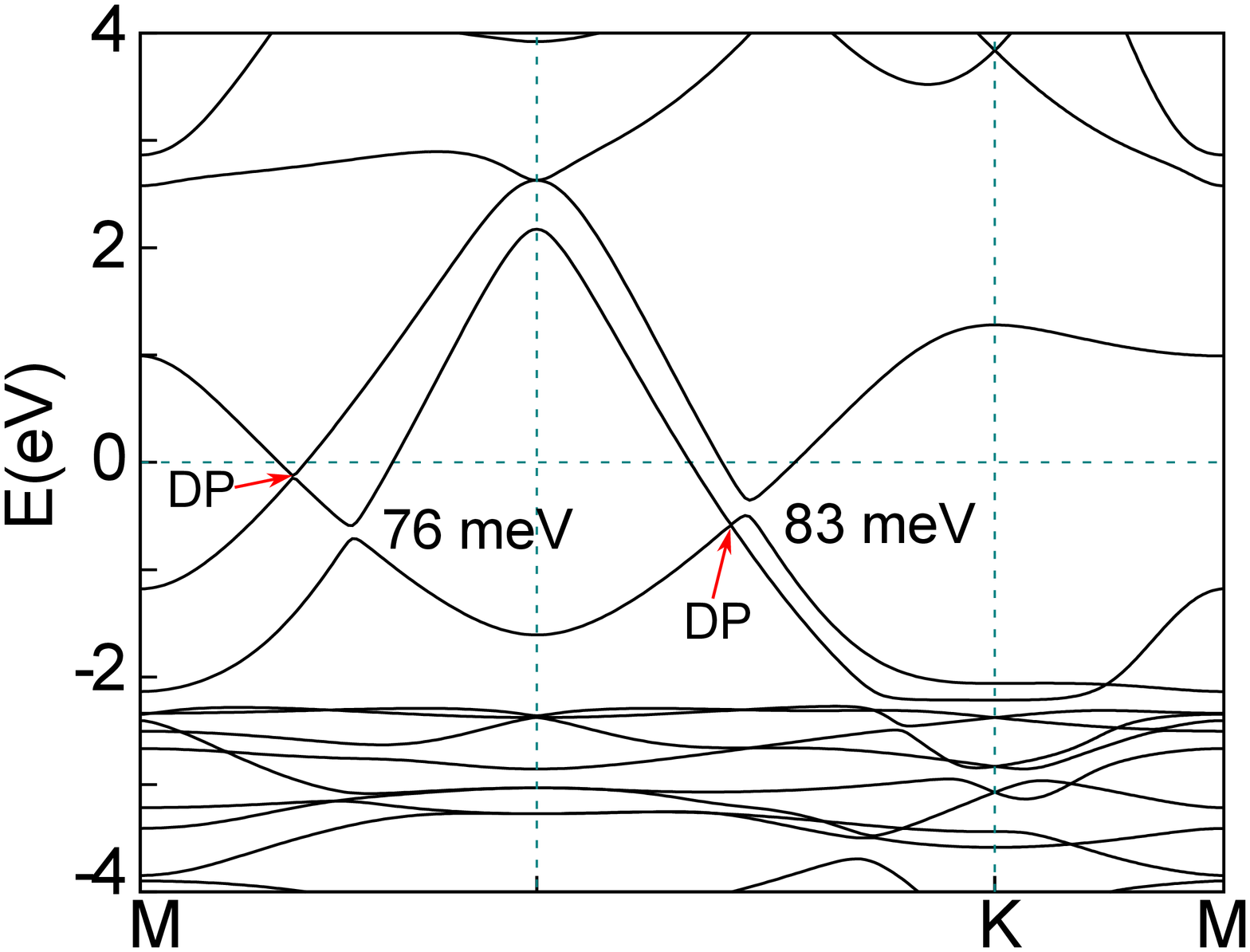}
\caption{ Calculated band structure of buckled Cu$_2$Ge monolayer where neighbouring Cu atoms are buckled with a height difference of 0.1 {\AA}. We find that the DNL containing A and A$^{\prime}$ (B and B$^{\prime}$) is transformed into three Dirac points (DPs) along the three nonequivalent ${\Gamma}-$K (${\Gamma}-$M) lines. These DPs are protected by three nonequivalent $M_{\rm \sigma}$ or $C_2$ symmetries, similar to the case of Cu$_2$Si monolayer~\cite{cu2si}.The sizes of gaps along the ${\Gamma}-$M and ${\Gamma}-$K lines are given.}
\end{figure}


\newpage

{\bf 5. Projected bands onto Fe and Ge/Sn orbitals.}
\begin{figure}[ht]
\includegraphics[width=16cm]{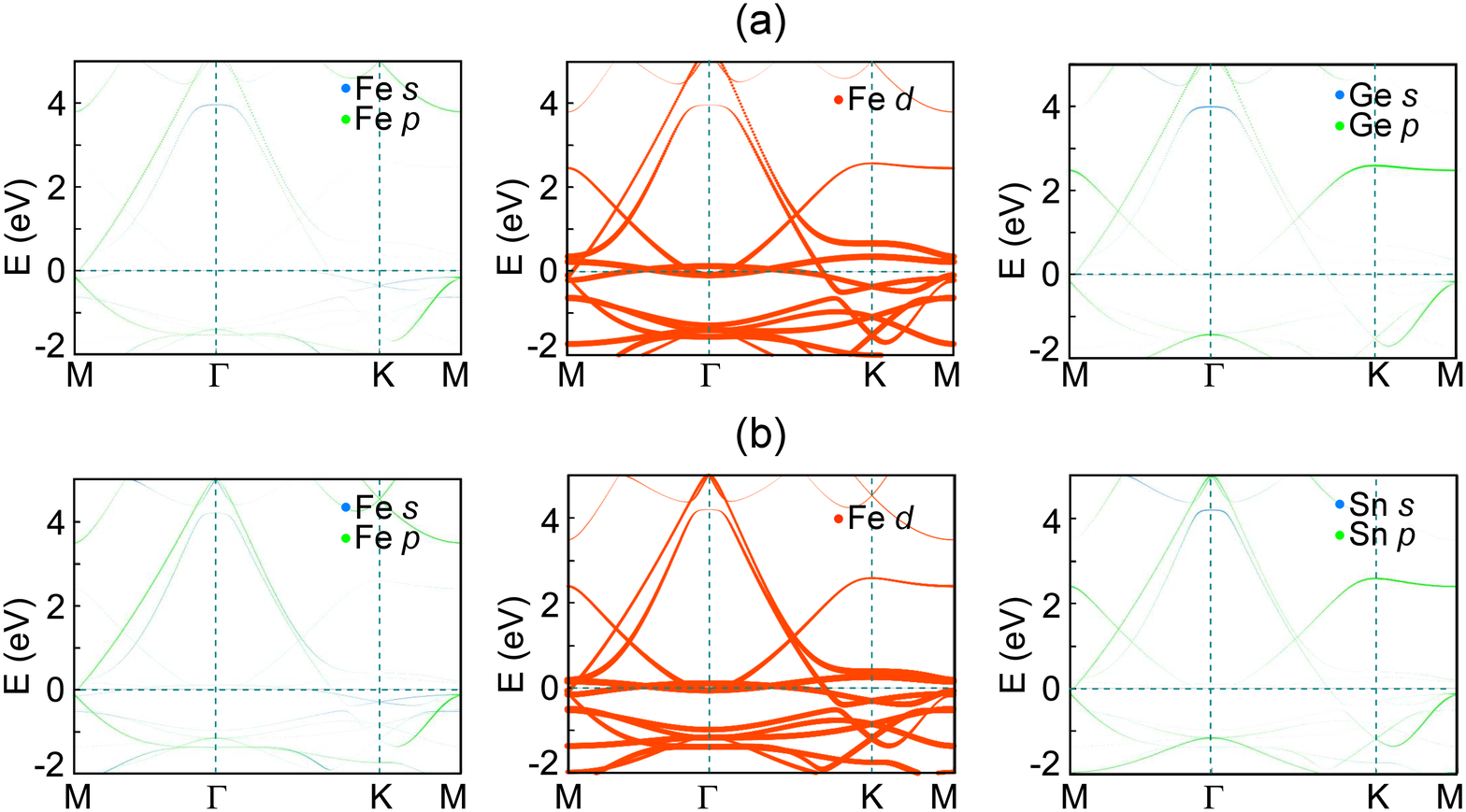}
\caption{ Calculated projected bands of the NM (a) Fe$_2$Ge and (b) Fe$_2$Sn monolayers onto the Fe 4$s$, 4$p$, 3$d$ orbitals, Ge 4$s$, 4$p$ orbitals, and Sn 5$s$, 5$p$ orbitals. Here, the radii of circles are proportional to the weights of the corresponding orbitals. The Fe 3$d$ orbitals are more dominant components of the electronic states around $E_{\rm F}$, compared to other orbitals.}
\end{figure}

\vspace{1.2cm}

{\bf 6. Stoner criteria for ferromagnetism in Fe$_2$Ge and Fe$_2$Sn monolayers.}
\begin{figure}[ht]
\includegraphics[width=16cm]{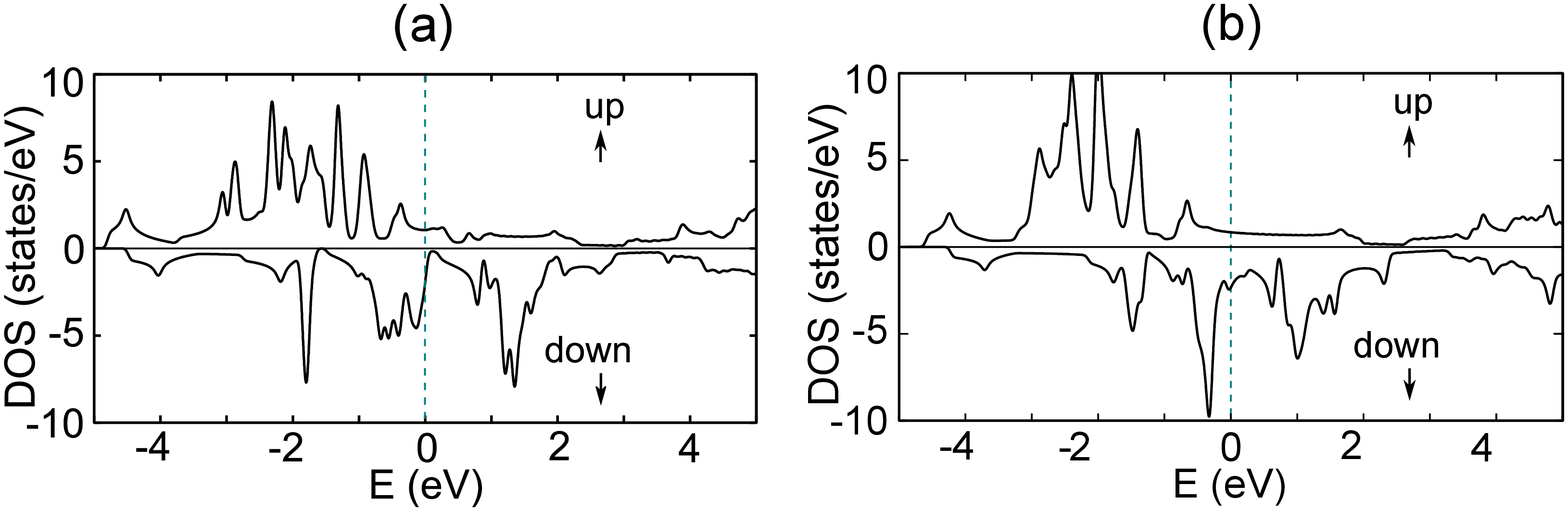}
\caption{ Calculated spin-polarized density of states (DOS) for the FM (a) Fe$_2$Ge and (b) Fe$_2$Sn monolayers. The Stoner parameter $I$ can be estimated with dividing the exchange splitting ${\Delta}E$ of spin-up and spin-down state density by the corresponding magnetic moment $m$. Here, we obtain ${\Delta}E$ = 1.152 eV for Fe$_2$Ge monolayer and ${\Delta}E$ = 1.503 eV for Fe$_2$Sn monolayer by calculating the average difference of the Kohn-Sham eigenvalues of spin-up and spin-down bands below Fermi level~\cite{stoner-sm1, stoner-sm2}. Using the relation ${\Delta}E$ = $I$·$m$, the Stoner parameter $I$ = ${\Delta}E$/$m$ is calculated to be 0.553 eV (0.632 eV) for Fe$_2$Ge (Fe$_2$Sn) monolayer. Meanwhile, from Fig. 3(a) [3(b)], the DOS of the NM Fe$_2$Ge (Fe$_2$Sn) monolayer is 4.32 state/eV (5.53 states/eV) per spin at the Fermi level. It is thus demonstrated that the Stoner’s criterion D($E_{\rm F}$)·$I$ $>$ 1 is satisfied as 2.388 (3.509) for Fe$_2$Ge (Fe$_2$Sn) monolayer.}
\end{figure}


\newpage

{\bf 7. Band structure of the FM Fe$_2$Ge monolayer.}
\begin{figure}[ht]
\includegraphics[width=11.5cm]{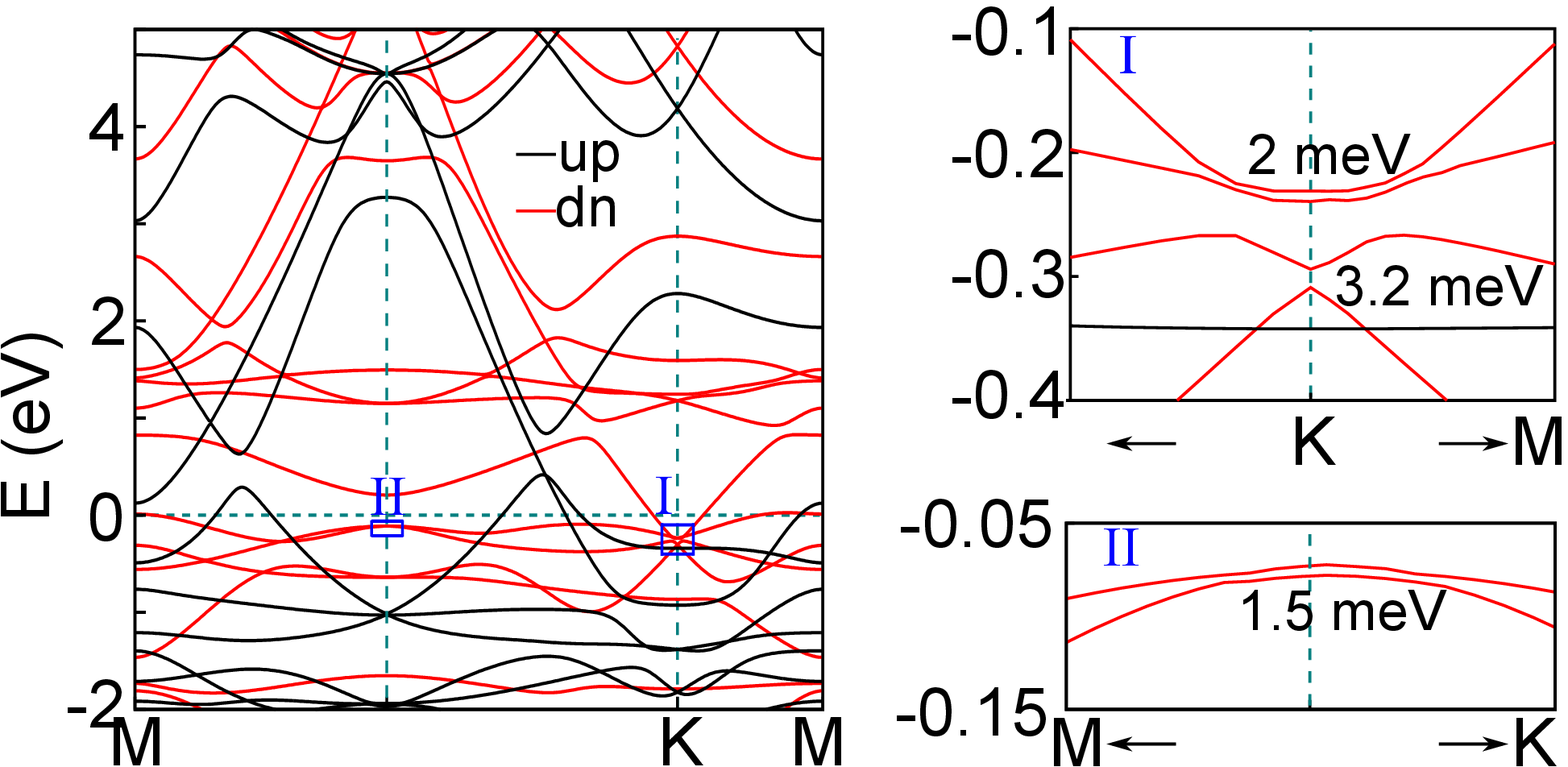}
\caption{ Calculated band structure of the FM Fe$_2$Ge monolayer in the absence of SOC and zoom-in band structures below $E_{\rm F}$. The numbers represent the gaps.}
\end{figure}

\vspace{1.2cm}

{\bf 8. Symmetry-protected 2D Weyl points in Fe$_2$Ge monolayer.}
\begin{figure}[ht]
\includegraphics[width=13.5cm]{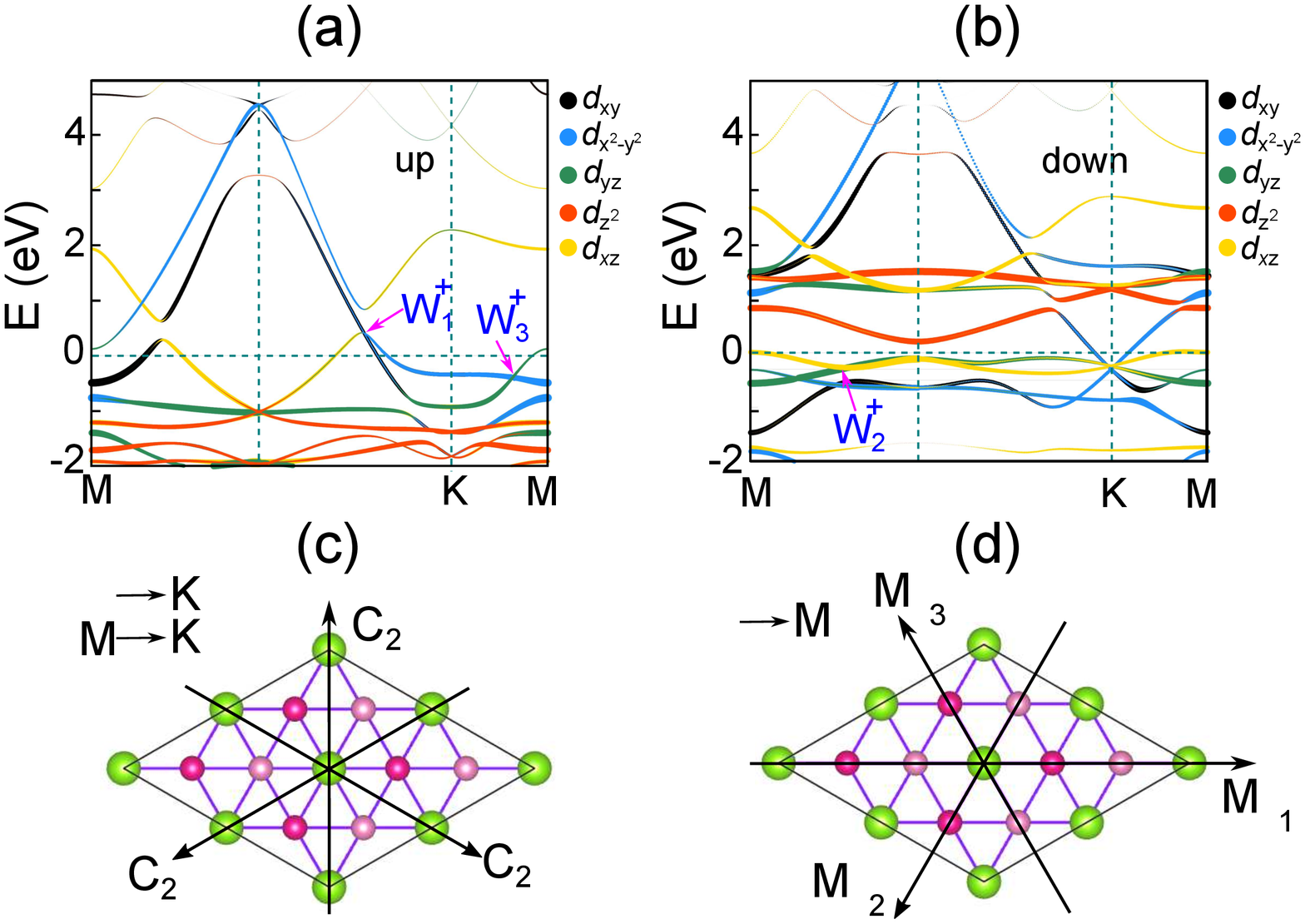}
\caption{ Projected (a) spin-up and (b) spin-down bands of Fe$_2$Ge monolayer onto the Fe 3$d$ orbitals, obtained using DFT calculations without SOC. Here, the radii of circles are proportional to the weights of the corresponding orbitals. In (a), we find that the two crossing bands at W$^{+}_{1}$  arise from the Fe \textit{d$_{xy}$} and \textit{d$_{xz}$} orbitals, whereas those at W$^{+}_{3}$ arise from the Fe \textit{d$_{yz}$} and \textit{d$_{x^2-y^2}$} orbitals. Note that, along the ${\Gamma}-$K and M$-$K directions, the little group has three nonequivalent $C_2$ rotation symmetries, as shown in (c). The two crossing bands at W$^{+}_{1}$  or W$^{+}_{3}$  located in the ${\Gamma}-$K and M$-$K lines have the opposite parity of $C_2$ rotation symmetry, thereby remaining gapless. In (b), we find that the two crossing bands at W$^{+}_{2}$  arise from the Fe \textit{d$_{xz}$}  and \textit{d$_{yz}$} orbitals. Note that, along the ${\Gamma}-$M direction, the little group has three out-of-plane mirror symmetries, $M_{\rm \sigma1}$, $M_{\rm \sigma2}$ and $M_{\rm \sigma3}$, as shown (d). The crossing bands at W$^{+}_{2}$  located in the ${\Gamma}-$M line have the opposite parity of $M_{\rm \sigma}$, thereby remaining gapless.}
\end{figure}

\newpage

{\bf 9. WNLs of Fe$_2$Sn monolayer in the absence of SOC.}
\begin{figure}[ht]
\includegraphics[width=12cm]{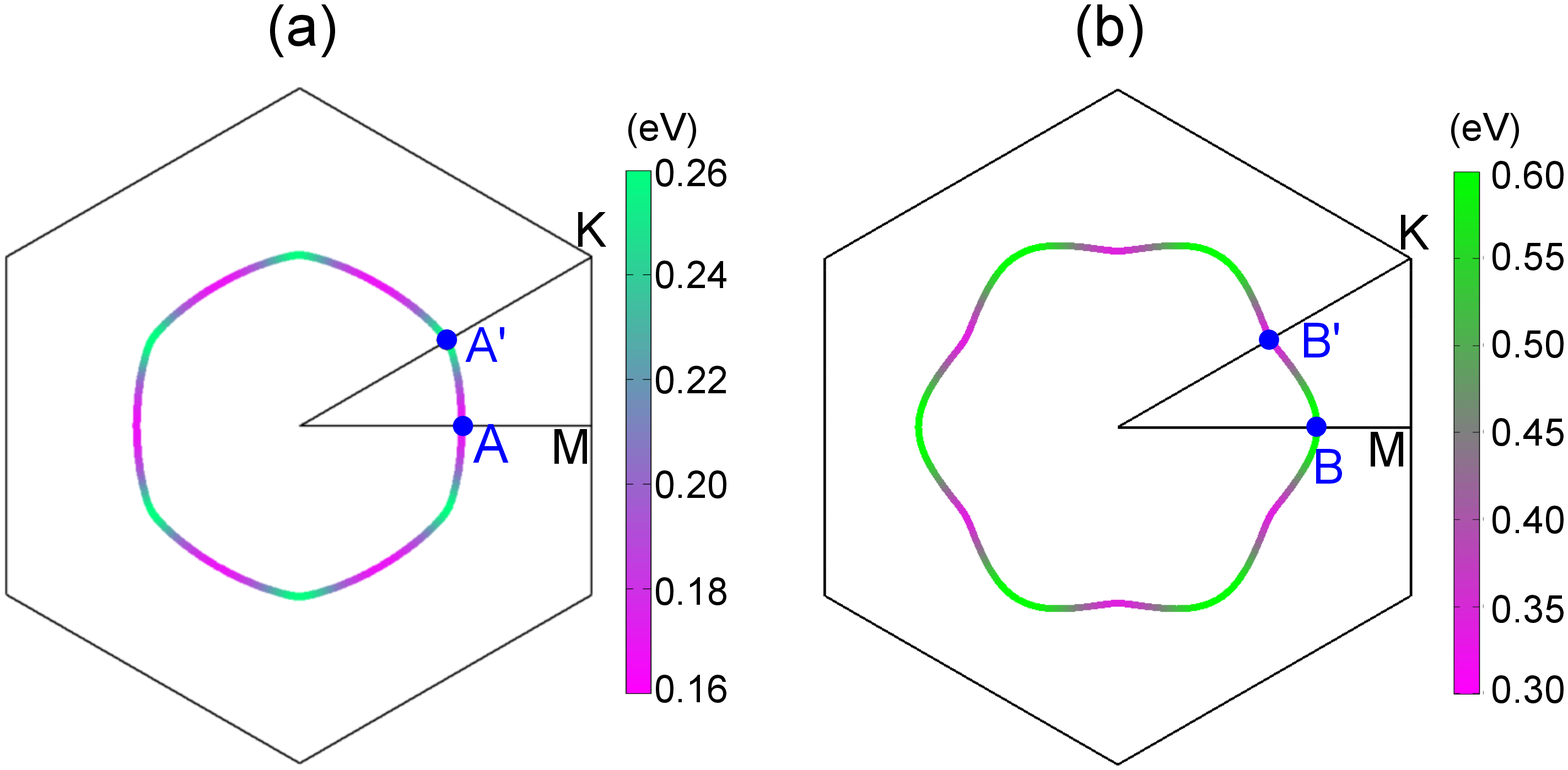}
\caption{ Momentum distribution of WNLs passing through the (a) A-A$^{\prime}$ and (b) B-B$^{\prime}$ points in Fe$_2$Sn monolayer, obtained using DFT calculations without SOC. The energy eigenvalues along the WNLs are drawn in the color scale.}
\end{figure}

\vspace{1.2cm}

{\bf 10. Projected bands of FM Fe$_2$Sn monolayer onto the Fe 3$d$ orbitals.}
\begin{figure}[ht]
\includegraphics[width=12.cm]{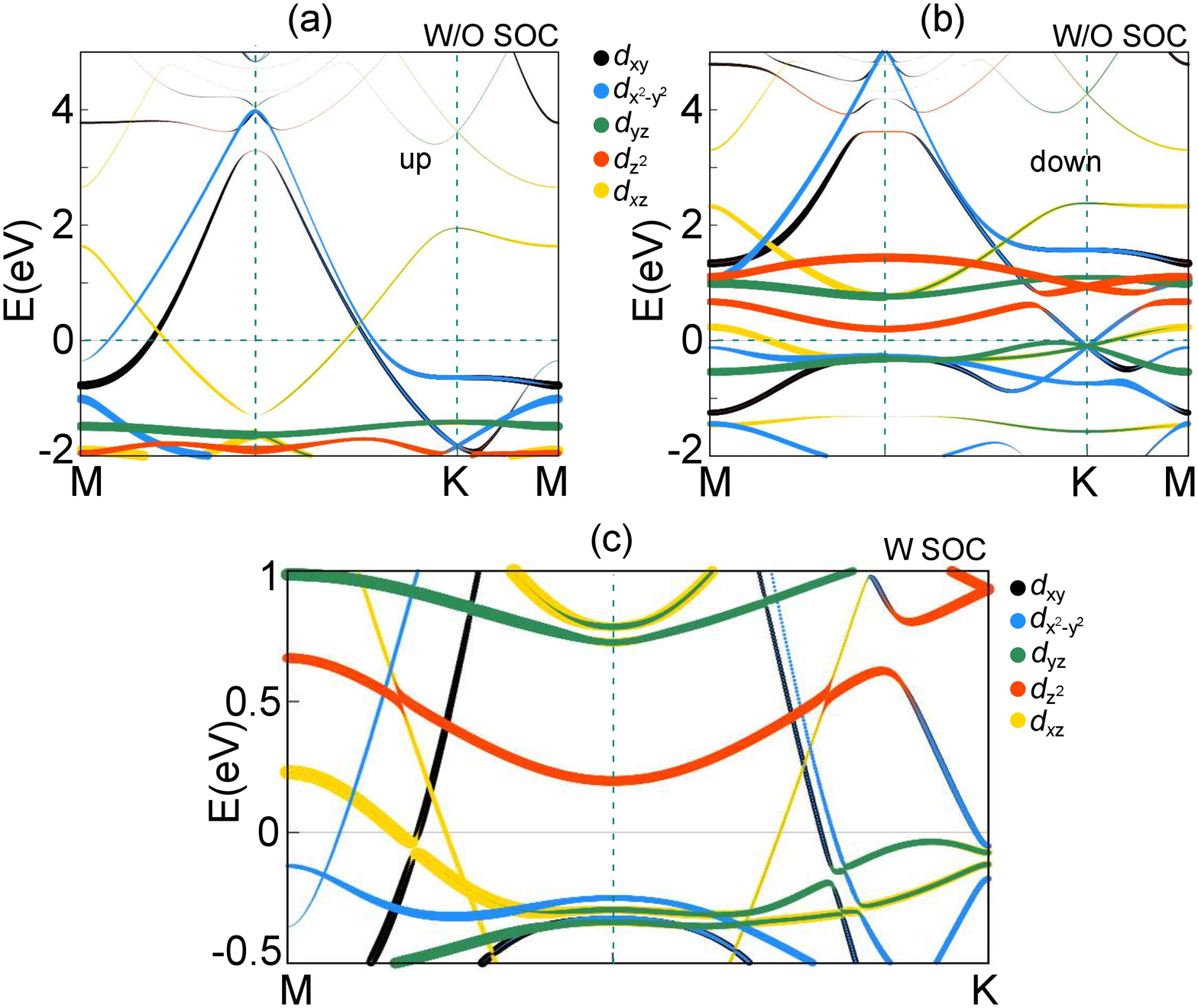}
\caption{ Projected (a) spin-up and (b) spin-down bands of Fe$_2$Sn monolayer onto the Fe 3$d$ orbitals, obtained using DFT calculations without SOC. Here, the radii of circles are proportional to the weights of the corresponding orbitals. In (a), we find that the $\alpha$ ($\beta$) band mainly arises from the Fe \textit{d$_{x^2-y^2}$} (\textit{d$_{xy}$}) orbital with the even parity of $M_{\rm z}$, while the $\gamma$ band is composed of the Fe \textit{d$_{xz}$} orbital with the odd parity of $M_{\rm z}$. The projected bands with including SOC are given in (c). Here, the $\chi$ band arising from the Fe \textit{d$_{z^2}$} orbital hybridizes with the $\gamma$ band because the two bands have same parity (+$i$) of $M_{\rm z}$. Meanwhile, the $\chi$ band does not hybridize with the $\alpha$ and $\beta$ bands having the opposite parity (-$i$) of $M_{\rm z}$, forming WNLs.}
\end{figure}

\newpage

{\bf 11. WNLs in Fe$_2$Sn monolayer in the presence of SOC.}
\begin{figure}[ht]
\includegraphics[width=11.cm]{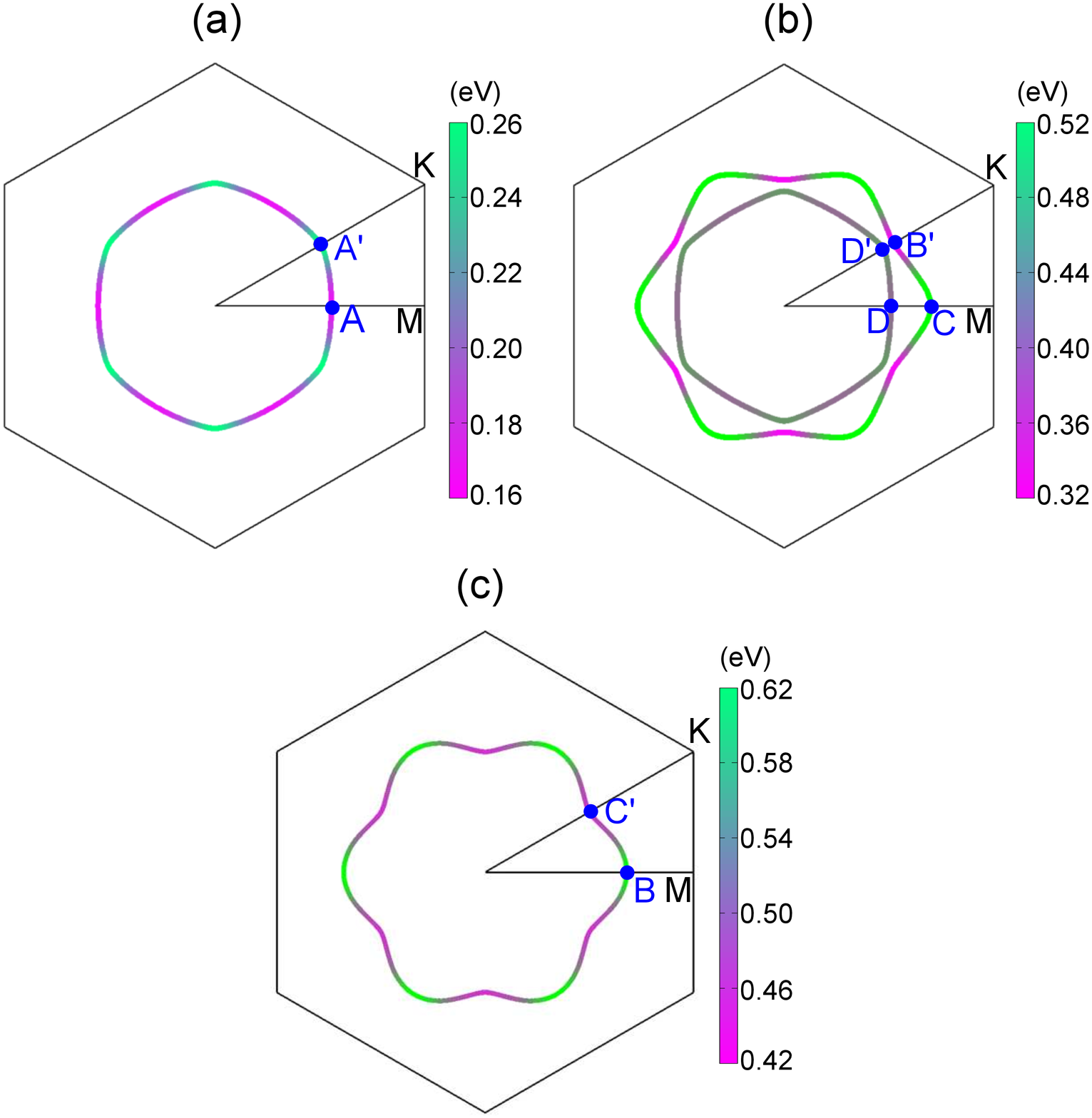}
\caption{ Momentum distribution of WNLs passing through the (a) A$-$A$^{\prime}$, (b) C$-$B$^{\prime}$ and D$-$D$^{\prime}$, and (c) B$-$C$^{\prime}$ points in Fe$_2$Sn monolayer, obtained using DFT calculations with SOC. The energy eigenvalues along the WNLs are drawn in the color scale.}
\end{figure}

\vspace{1.2cm}

{\bf Table S1: Calculated lattice constants for various monolayers. For Fe$_2$Ge and Fe$_2$Sn, the values for the NM and FM phases are given.}
\begin{table}[ht]
\centering
\begin{ruledtabular}
\begin{tabular}{lccccc}
    & NM Cu$_2$Ge & NM Fe$_2$Ge & FM Fe$_2$Ge & NM Fe$_2$Sn & FM Fe$_2$Sn \\  \hline
a=b ({\AA}) & 4.218 & 4.063 & 4.147 & 4.306 & 4.453
\end{tabular}
\end{ruledtabular}
\end{table}

\end{flushleft}

\end{document}